\documentclass[iop]{emulateapj}

\usepackage{amsmath}

\begin{document}

\title {Electron-impact excitation of Fe$^{2+}$: a comparison of intermediate coupling 
frame transformation, Breit-Pauli and Dirac $R$-matrix calculations }
\shorttitle{Electron-impact excitation of Fe$^{2+}$}
\shortauthors{Badnell and Ballance}

\author{N.~R. Badnell$^1$ and C.~P. Ballance$^2$} 
\affil{$^1$ Department of Physics, University of Strathclyde, Glasgow, G4 0NG, UK}
\affil{$^2$ Department of Physics, Auburn University, Auburn, AL 36849-5311, USA}

\begin{abstract}
Modeling the spectral emission of low-charge iron group ions enables the diagnostic
determination of the local physical conditions of many cool plasma environments such as 
those found in H {\sc ii} regions, planetary nebulae, active galactic nuclei etc. 
Electron-impact excitation drives the population of the emitting levels and, hence, 
their emissivities.
By carrying-out Breit-Pauli and intermediate coupling frame transformation (ICFT) 
$R$-matrix calculations for the electron-impact excitation of Fe$^{2+}$ 
which both use the exact same atomic structure and the same close-coupling expansion, 
we demonstrate the validity of the application of the powerful ICFT method to low-charge
iron group ions. This is in contradiction to the finding of Bautista et al. 
[\apjl, 718, L189, (2010)]  who carried-out ICFT and Dirac $R$-matrix calculations 
for the same ion. We discuss possible reasons. 

\end{abstract}


\section{Introduction}
\label{sec:intro}



\begin{table*}
\caption{Lowest 17 energy levels (Ryd) of the $3d^6$ ground configuration of Fe$^{2+}$. 
\label{tab:energies}}
\begin{small}
\begin{center}
\begin{tabular}{lllll}
        \tableline
	\tableline
 Term/Level & Observed\tablenotemark{a} & AS\tablenotemark{b} & GRASP0\tablenotemark{b}  & GRASP0\tablenotemark{c}  \\
        \tableline
 $^5D_4$ &   0.0        & 0.0      & 0.0       & 0.0      \\
 $^5D_3$ &   0.003975   & 0.004001 & 0.003408  & 0.003613 \\
 $^5D_2$ &   0.006733   & 0.006803 & 0.005823  & 0.006162 \\
 $^5D_1$ &   0.008497   & 0.008602 & 0.007381  & 0.007803 \\
 $^5D_0$ &   0.009361   & 0.009483 & 0.008147  & 0.008608 \\
 $^3P2_2$&   0.176830   & 0.18434  & 0.19477   & 0.19564  \\
 $^3P2_1$&   0.188527   & 0.19636  & 0.20543   & 0.20690  \\
 $^3P2_0$&   0.193266   & 0.20122  & 0.20994   & 0.21161  \\
 $^3H_6$ &   0.182719   & 0.20976  & 0.22119   & 0.22244  \\
 $^3H_5$ &   0.184995   & 0.21173  & 0.22282   & 0.22416  \\
 $^3H_4$ &   0.186645   & 0.21370  & 0.22444   & 0.22588  \\
 $^3F2_4$&   0.195578   & 0.21026  & 0.22199   & 0.22313  \\
 $^3F2_3$&   0.197744   & 0.21313  & 0.22439   & 0.22571  \\
 $^3F2_2$&   0.199177   & 0.21468  & 0.22572   & 0.22712  \\
 $^3G_5$ &   0.223796   & 0.24873  & 0.26285   & 0.26437  \\
 $^3G_4$ &   0.227278   & 0.25202  & 0.26558   & 0.26729  \\
 $^3G_3$ &   0.229114   & 0.25367  & 0.26700   & 0.26878  \\
\tableline
\tablenotetext{1}{\cite{NIST}}
\tablenotetext{2}{Present work.}
\tablenotetext{3}{From \cite{Bau10}}

\end{tabular}
\end{center}
\end{small}
\end{table*}


The electron-impact excitation of low-charge ion group ions is an important process
in many cool astrophysical plasmas. The subsequent spectral emission can be used to diagnose
the physical conditions of H {\sc ii} regions, planetary nebulae, active galactic nuclei and
many stars --- see for example \cite{Joh00,Ves01,Est04,Mes09,Meh10,Mad12,Zha12,Mcc13}.

In order to have confidence in the temperatures, densities, and chemical compositions deduced
from modeling such emission it is important that the uncertainties in the atomic
data be well understood and their effects modeled. Energy levels can be compared
with observed. Radiative data can usually be computed with increasingly large
configuration interaction expansions to demonstrate convergence to a sufficient degree.
The case of electron-impact excitation is more problematic. The complexity of
the open $3d$-subshell leads to a large number of low-lying target levels 
which should be coupled together in a scattering calculation \citep{Burk94}.
The $R$-matrix method is paramount here. It has been advantageous to introduce
some approximations into the scattering problem which leads to the highly efficient 
and widely used intermediate coupling frame transformation (ICFT) $R$-matrix method
\citep{Griff98}.

\cite{Bau10} recently modeled Fe {\sc iii} line intensities from the Orion nebulae which
had been observed by \cite{Mes09}. They found much better accord between theory and
observation when they used excitation data which they calculated with the Dirac Atomic 
$R$-matrix code (DARC) methodology \citep{Ait96, Nor04} than that they calculated with the 
ICFT method. Their ICFT Maxwellian-averaged collision strengths
for transitions from the ground level to all other levels of the $3d^6$ configuration
differed by up to a factor 3 (both larger and smaller) from their DARC ones at 
$10^{4}$~K.
They attributed this difference ``to the limited treatment of spin-orbit effects in
near-threshold resonances'' by the ICFT method. This surprising finding has
implications not just for the validity and use of the ICFT method but also the 
$R$-matrix II \citep{Burk94} FINE method \citep{Burk10} with which it shares some similarities.

There are a number of issues which can be raised about the comparison made by 
\cite{Bau10}.  It has long been known that to be able to deduce
anything meaningful about two scattering methods from a comparison of collision data
it is necessary that the differences in the atomic structure be kept to an absolute
minimum. \cite{Bau10}  used a much larger configuration interaction (36 configurations)
for their ICFT atomic structure compared to that for their DARC one (8 configurations). 
They also used a number of pseudo orbitals in their ICFT target description while the DARC
ones were purely spectroscopic --- pseudo-orbitals lead to pseudo-resonances.
They retained 283 levels in their ICFT close-coupling expansion but 322 levels in their
DARC one. 

We report-on the results of parallel ICFT and BPRM $R$-matrix calculations for Fe$^{2+}$ that use
the exact same atomic structure (an 8 configuration interaction expansion) and the same
close-coupling expansion (322 levels). We find excellent agreement between the two ($<5$\% 
where \cite{Bau10}  found factors of 3). We also find excellent agreement with a similar
DARC calculation which we have carried out. Many of the differences found by
\cite{Bau10}  disappear once we correctly label their DARC transitions.
Some differences remain with their ICFT results and for which the source is likely
one of the possible causes listed in the preceding paragraph. 

The rest of this paper is structured as follows: in Section 2 we describe the key
differences between the different $R$-matrix methods we use; in Section 3 we give
details of both the atomic structure and atomic collision calculations; in Section 4
we present our results and discuss their agreement with and differences from the
results of \cite{Bau10}; and in Section 5 we draw some key conclusions.

\section{Theory}
\label{sec:theory}

Relativistic effects can be expected to be important in general for the iron group 
--- just compare energies and radiative data calculated with and without such an allowance. 
The case for their inclusion in the description of electron-impact excitation at low stages 
of ionization is less clear. The Iron Project argued \citep{Hum93} that algebraic recoupling
of $LS$-coupling scattering matrices alone was sufficient to determine level-resolved collision
data and adopted such an approach. Essentially this is because electron-impact excitation
is mediated by a two-body operator while the dominant fine-structure operator is one-body.
Only more recently \citep{Ram07} has it become practical to carry-out
similarly sized $R$-matrix calculations which include relativistic interactions explicitly.
Large differences with the earlier Iron Project results have been noted \citep{Ram07, Bau10}
but it is difficult to separate out the difference due to the choice of
the collision representation from the difference due to the differing
atomic structure representations adopted. Obtaining a sufficiently accurate
atomic structure is itself a non-trivial exercise for these ions.

Relativistic interactions can be included using either the Breit-Pauli (BPRM) or Dirac 
Atomic (DARC) $R$-matrix codes. A full Dirac treatment is not necessary here ($Z\le 30$)
but the DARC suite of codes has lent itself more readily to large-scale parallelization.
Level-resolved $R$-matrix inner-region codes lead to the need to diagonalize large 
Hamiltonians (e.g. rank $>10^5$) and to solve a large set of coupled equations
(e.g. $\sim 10000$) in the $R$-matrix outer region. 

The BPRM method lends itself to useful approximation because
the scattering problem is set-up initially in term-resolved $LS$-coupling and subsequently
transformed to level-resolved $jK$-coupling upon the addition of fine-structure
relativistic operators (e.g. spin-orbit). If the relativistic operators are not
too large then they can be treated perturbatively. The $R$-matrix II approach \citep{Burk94}
diagonalizes the $LS$-coupling scattering Hamiltonian and then transforms the
resulting eigen-vectors (surface amplitudes) to $jK$-coupling. It allows for
relativistic effects through a further transformation which makes use of term coupling 
coefficients \citep{Jon75}.
The outer region problem is then solved as in the original BPRM method.
This reduces the computational time (for the diagonalization) by an order of 
magnitude or allows a much larger scattering expansion to be used instead. 

The intermediate coupling frame transformation (ICFT) $R$-matrix method \citep{Griff98}
goes a step further. It solves the outer region coupled channel problem
in $LS$-coupling as well. It then transforms (asymptotically) the 
scattering or reactance matrices to $jK$-coupling and again uses term 
coupling coefficients. The ICFT method is an order of magnitude 
more efficient in the outer region as well. It does require that it is
sufficiently accurate to delay the treatment of fine-structure relativistic
effects until reaching asymptopia\footnote{All BPRM/ICFT methods treat 
the one-body non-fine-structure operators without further approximation --- they
are included in the scattering Hamiltonian to be diagonalized.}.
The original comparisons \citep{Griff98, Griff99, Bad99} between the results of ICFT and BPRM 
calculations demonstrated agreement between the Maxwellian-averaged collision strengths to 
within a few percent. This is well within the uncertainties
for complex ions which are introduced by truncating both the target configuration
interaction expansion and the close-coupling scattering expansion.


\begin{table*}
\caption{Maxwellian effective collision strengths for  Fe$^{2+}$ at $10^4$~K\\
from the $^5D_4$ ground-level  to the next 16 levels as listed in Table~\ref{tab:energies}.
\label{tab:upsilons}}
\begin{small}
\begin{center}
\begin{tabular}{llllllll}
        \tableline
	\tableline
 Term/Level & LS(jK)J\tablenotemark{a} & ICFT(Obs.)\tablenotemark{a} & ICFT\tablenotemark{a} & BPRM\tablenotemark{a} & 
DARC\tablenotemark{a} & DARC\tablenotemark{b} & ICFT\tablenotemark{b} \\
        \tableline
 $^5D_3$ & 3.19(+0)   & 3.09(+0)\tablenotemark{c}   & 3.11(+0)   & 3.03(+0)   & 2.73(+0)    & 2.54(+0)     & 4.57(+0)    \\
 $^5D_2$ & 1.45(+0)   & 1.45(+0)   & 1.45(+0)   & 1.41(+0)   & 1.22(+0)    & 1.11(+0)     & 1.94(+0)    \\
 $^5D_1$ & 6.79($-$1) & 7.14($-$1) & 6.85($-$1) & 6.65($-$1) & 5.91($-$1)  & 5.33($-$1)   & 8.79($-$1)  \\
 $^5D_0$ & 1.99($-$1) & 2.14($-$1) & 2.01($-$1) & 1.97($-$1) & 1.74($-$1)  & 1.60($-$1)   & 2.51($-$1)  \\
 $^3P2_2$& 6.99($-$1) & 7.13($-$1) & 7.94($-$1) & 8.06($-$1) & 7.00($-$1)  & 7.14($-$1)   & 7.14($-$1)  \\
 $^3P2_1$& 2.16($-$1) & 1.79($-$1) & 2.22($-$1) & 2.20($-$1) & 1.80($-$1)  & 1.96($-$1)   & 1.84($-$1)  \\
 $^3P2_0$& 3.39($-$2) & 3.07($-$2) & 4.35($-$2) & 3.79($-$2) & 3.05($-$2)  & 3.25($-$2)   & 3.83($-$2)  \\
 $^3H_6$ & 1.41(+0)   & 1.45(+0)   & 1.43(+0)   & 1.44(+0)   & 1.30(+0)    & 1.21(+0)     & 2.66(+0)    \\
 $^3H_5$ & 5.05($-$1) & 6.44($-$1) & 6.38($-$1) & 6.62($-$1) & 5.50($-$1)  & 5.33($-$1)   & 1.10(+0)  \\
 $^3H_4$ & 9.45($-$2) & 2.48($-$1) & 2.16($-$1) & 2.20($-$1) & 2.11($-$1)  & 1.91($-$1)   & 2.41($-$1)  \\
 $^3F2_4$& 1.10(+0)   & 1.16(+0)   & 1.12(+0)   & 1.13(+0)   & 9.96($-$1)  & 9.84($-$1)   & 1.47(+0)  \\
 $^3F2_3$& 4.51($-$1) & 5.43($-$1) & 5.15($-$1) & 5.26($-$1) & 4.52($-$1)  & 4.54($-$1)   & 6.42($-$1)  \\
 $^3F2_2$& 1.60($-$1) & 1.99($-$1) & 1.88($-$1) & 1.95($-$1) & 1.66($-$1)  & 1.73($-$1)   & 2.11($-$1)  \\
 $^3G_5$ & 1.34(+0)   & 1.27(+0)   & 1.31(+0)   & 1.32(+0)   & 1.07(+0)    & 1.11(+0)     & 1.11(+0)    \\
 $^3G_4$ & 5.29($-$1) & 4.90($-$1) & 5.36($-$1) & 5.51($-$1) & 4.15($-$1)  & 4.21($-$1)     & 1.24(+0)    \\
 $^3G_3$ & 1.28($-$1) & 1.40($-$1) & 1.62($-$1) & 1.67($-$1) & 1.21($-$1)  & 1.35($-$1)   & 4.52($-$1)  \\
        \tableline
\tablenotetext{1}{Present work.}
\tablenotetext{2}{\cite{Bau10} (but DARC order corrected.)}
\tablenotetext{3}{(m) denotes $\times 10^m$.}
\end{tabular}
\end{center}
\end{small}
\end{table*}


\section{Calculational Details}
\label{sec:calcs}

\subsection{Structure}

We used the same 8 configuration interaction target expansion for our ICFT, BPRM and DARC
calculations: $3s^2 3p^6 3d^6$, $3s^2 3p^6 3d^5 4s$, $3s^2 3p^6 3d^5 4p$, $3s^2 3p^5 3d^7$,
$3s^2 3p^4 3d^8$, $3s^2 3p^4 3d^7 4s$, $3s^2 3p^4 3d^7 4p$, and $3p^6 3d^8$. This is the same
target expansion that was used by \cite{Bau10}  in their DARC calculation. It gives rise to
994 terms and 2578 levels. All orbitals are taken to be physical ones. 

We used the program {\sc autostructure} \citep{Bad11} to generate our common ICFT/BPRM target.
The scaling parameters associated with the Thomas-Fermi-Dirac-Amaldi model potentials 
were determined by an iterative variational scheme. They are given by $\lambda_{1s-3p}=1.110$,
$\lambda_{3d}=1.024$, $\lambda_{4s}=1.002$, and $\lambda_{4p}=1.180$. 

Our DARC target was generated using GRASP0 \citep{Par96, Nor04} by varying all orbitals 
simultaneously. The GRASP0 (and subsequent DARC)
calculations were `repeated' because it was initially thought that the original 
detailed results (e.g. collision strengths) were lost. They were recovered subsequently.
We do not have access to the level energies for the structure which \cite{Bau10}  used
for their ICFT collision calculation.

We compare our {\sc autostructure} energies (AS) for the lowest 17 levels with those from
GRASP0 in Table~\ref{tab:energies}. The two sets of GRASP0 energies differ slightly.
The original orbitals were optimized slightly differently it seems.
This provides us with an opportunity to look at the sensitivity of the subsequent
DARC collision strengths to such target differences. We compare our calculated energies
with those obtained from \cite{NIST} as well. They lie 10--15\% above the observed ones
for levels belonging to excited terms. Those from {\sc autostructure} are slightly closer
to observed. The {\sc autostructure}/GRASP0 energies lie above/below the observed for
levels of the ground term. The {\sc autostructure} ones are markedly closer ($\sim 1\%$)
to observed than the GRASP0s' ($10-15\%$). 

\subsection{Scattering}

We used the same 3 configuration close-coupling target expansion for our ICFT, BPRM and DARC
calculations: $3s^2 3p^6 3d^6$, $3s^2 3p^6 3d^5 4s$, and $3s^2 3p^6 3d^5 4p$.
It gives rise to 136 terms and 322 levels. There is nearly a 2 Ryd gap to the next 
(configuration interaction) level (323).
We included all partial waves explicitly up to $2J=59$. The contribution from higher-$J$
was `topped-up' following the procedures of \cite{Burg74} for dipole transitions and \cite{Bad01} 
for non-dipole allowed. The ICFT/BPRM calculations explicitly dropped electron exchange
above $2J=19$ for efficiency.  We used 11 continuum basis orbitals and our scattering
energy extended to 4 Ryd. We used an energy step of $5\times 10^{-5} z^2\quad (z=2) $ Ryd 
in the resonance region leading to 10000 energies. We used a step of $10^{-3} z^2$ Ryd
elsewise. The Maxwellian convolution of the ordinary collision strengths utilized their
infinite energy Born and Bethe limit points \citep{Burg92} to interpolate values at higher
energies as they were needed.

\section{Collision Results}
\label{sec:results}


\begin{figure}
\includegraphics[width=2.75in, angle=-90]{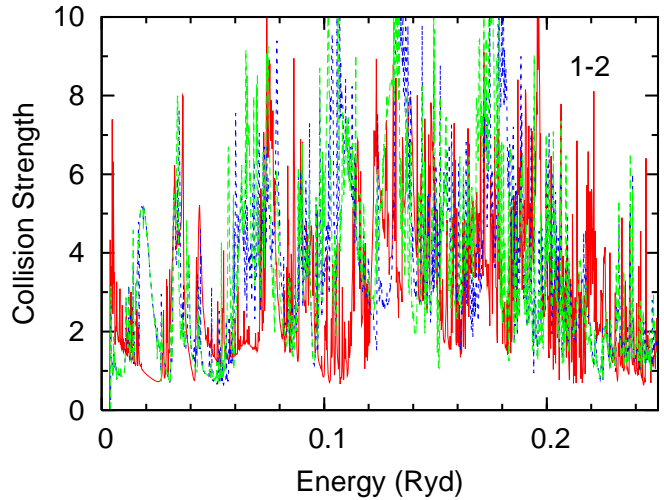}
\caption{Collision strengths for the  $3d^6~^5D_4 - ^5D_3$ transition in Fe$^{2+}$: 
ICFT (green long-dashed curve) and Breit-Pauli (blue short-dashed curve) $R$-matrix from the present work; 
Dirac $R$-matrix (red solid curve) due to \cite{Bau10}.}
\label{fig:omegas}
\end{figure}


In Table~\ref{tab:upsilons} we present and compare effective collision strengths at
$10^4$~K for transitions from the Fe$^{2+}$ $^5D_4$ ground level to the next 16 levels.
The most immediate and important observation is the very close agreement between our
present ICFT and BPRM results ($<5\%$). Only the relatively weak excitation of the $^3P2_0$
level differs by more (13\%). We note next that the agreement between BPRM and the present
DARC results is typically 20\%. This can only be attributable to differences in the
atomic structure and is quite reasonable for this system --- \cite{Bau10}  discuss
the wide variation found in the literature. 

Where does that leave us with respect to the findings of \cite{Bau10}? If we compare 
our present DARC results with those that they give in their Table 1 we find complete
disagreement from the $^3H_5$ level on upwards and by factors that are comparable with
the differences they observed from their ICFT results. Investigation of their original 
detailed collision strengths reveals that the Maxwellian effective collision
strengths which they list in their Table~1 are in their DARC energy order.
If we look at the energies in our Table~\ref{tab:energies}
we see that the level ordering in their Table 1 (which we repeat in our Table
\ref{tab:upsilons}) does not correspond to energy order. Indeed, the result
which they attribute to $3d^6$~ $^3G_5$ actually belongs to their $3d^5 4s ^7S_3$ level.
We present their correctly ordered/labeled DARC results in our Table~\ref{tab:upsilons}.
They all agree with our present DARC results to within 10\%. This is consistent
with the degree of agreement of the target level energies from the two GRASP0
calculations which is much closer than with those from {\sc autostructure}.


\begin{figure}
\includegraphics[width=2.75in, angle=-90]{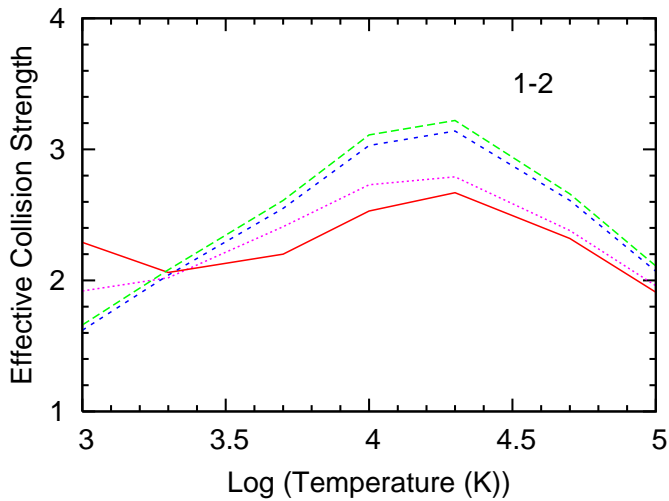}
\caption{Effective collision strengths for the  $3d^6~^5D_4 - ^5D_3$ transition in Fe$^{2+}$: 
ICFT (green long-dashed curve),  Breit-Pauli (blue short-dashed curve) and Dirac (purple dotted curve) 
$R$-matrix from the present work;  Dirac  $R$-matrix (red solid curve) due to \cite{Bau10}.}
\label{fig:upsilons}
\end{figure}


We turn next to the ICFT results of \cite{Bau10}  and which we show in our 
Table~\ref{tab:upsilons}. Most of the large factor differences from DARC
(and ICFT \& BPRM) have now disappeared except for the last two levels
which appear `reversed'. We do not have access to their original ICFT
results to check whether the labeling is correct or not. Remaining transitions
for which the original ICFT results are on the large side also show the
present ICFT/BPRM results to be 20--30\% larger than the original DARC ones.
Nevertheless, several largish differences remain with the $^3H_6$ one approaching
a factor of 2. We can only attribute this to their use of pseudo orbitals.
Our present ICFT and BPRM results differ by less than 1\% for this excitation.

\cite{Bau10}  noted good agreement between their ICFT and DARC background
collision strengths and attributed the cause of the difference in the
effective collision strengths to be due to the ICFT resonance structure being 
shifted to slightly lower energies --- see for example their Figure~2 
for the $3d^6~^5D_4 - ^5D_3$ transition.
By subsequent use of both sets of effective collision strengths to predict
line intensities measured in the Orion Nebula \cite{Bau10}  concluded that the DARC
ones were to be favored.

We compare our ICFT and BPRM collision strengths with the DARC ones of \cite{Bau10}
for this $(3d^6~^5D_4 - ^5D_3)$ 
transition in  Figure~\ref{fig:omegas}. We focus on the energy range $0-0.25$~Ryd 
since this covers the main resonance structure and $10^4$~K corresponds to $0.06$~Ryd.
There is no discernible shift in the  position of the ICFT and BPRM
resonance structures and they are in good overall agreement. The DARC collision 
strengths of \cite{Bau10} show a little weaker resonance structure compared to the
ICFT/BPRM ones, in so much as one can isolate it from the background. They are also 
somewhat different qualitatively. The present DARC results (not shown) are qualitatively 
very similar to the original DARC ones of \cite{Bau10}, but with a somewhat stronger 
resonance structure.  These agreements and differences reflect those observed in row one of
Table~\ref{tab:upsilons} for said methods.

In Figure~\ref{fig:upsilons} we compare effective collision strengths for 
the $3d^6~^5D_4 - ^5D_3$ transition over the temperature range $10^3 - 10^5$~K.
We see that $10^4$~K is representative of the largest differences. Our ICFT and BPRM 
effective collision strengths track each other over the entire temperature range.
Our present DARC ones are consistently closer to them than the original DARC
ones of \cite{Bau10}. The peak effective collision strength at $2\times 10^4$~K
reflects the maximal contribution of the strong resonance structure spread over $0-0.25$~Ryd.

We have investigated further the sensitivity of the effective collision strengths to 
`small' changes in resonance positions due to shifts in the threshold to which 
they are attached. We repeated our ICFT $R$-matrix calculation utilizing  observed 
energies for the lowest 127 levels \citep{NIST}. This includes all levels of the 
ground ($3s^2 3p^6 3d^6$) and first excited ($3s^2 3p^6 3d^5 4s$) configurations
as well as those levels of the $3s^2 3p^6 3d^5 4p$ configuration which overlap them.
Three levels within this range are not observed apparently ($3d^6$~$^1D_2$, $^1S_0$ and 
$3d^5 4s ^1S_0$). We shift them by amounts comparable to those used for similar
observed levels.

The largest changes to the effective collision strengths at $10^4$~K are typically
$\pm 15\%$  but most differences with our unshifted ICFT results are less than
$\pm 10\%$.  The one exception again is the $^3P2_0$ for which the shifted ICFT result
is 30\% smaller than the unshifted. We recall that the unshifted ICFT result
was 13\% larger than the BPRM. There is no particular systematic increase and/or 
decrease between the shifted and unshifted ICFT results. The differences of the 
GRASP0 energies from observed are larger than our {\sc autostructure} ones and 
so one might expect somewhat larger changes/uncertainties in the DARC effective
collision strengths as a result.

We close with the observation that if we neglect fine-structure relativistic
effects completely then our resulting pure algebraic $LS(jK)J$ recoupling
effective collision strengths agree with our ICFT ones to within 20\% except
for excitation of the $^3H_4$ level for which the difference is a factor of 2.
They are also in good accord with similar algebraic recoupling results of \cite{Zha96}
including for the $^3H_4$ now (not shown). 
There is no need for us to use such an approach because we have never reached
the stage of having carried-out the $LS(jK)J$ recoupling and not been able to 
carry-out the final term coupling transformation to take account of 
fine-structure effects. We note that the Iron Project also concluded that
algebraic recoupling sufficed for low-charge iron-group ions in most instances \citep{Hum93}.
With but a little extra effort we can effectively replace a full Breit-Pauli
$R$-matrix calculation by an $LS$-coupling one.

\section{Conclusions}
We have carried-out ICFT and Breit-Pauli\footnote{A full Breit-Pauli dataset for all 
322 levels (energies, radiative transition probabilities and effective collision strengths) 
is available in electronic form from the APAP website (\url{www.apap-network.org}).}
$R$-matrix calculations for the electron-impact excitation of Fe$^{2+}$ which use the
exact same atomic structure and the same close-coupling level expansion.  The results
demonstrate that the ICFT $R$-matrix method can be expected to provide accurate effective 
collision strengths for near neutral iron group ions which are well within the uncertainties
which exist due to the accuracy of the representation of their atomic structure.

The advantage of the ICFT approach over the Breit-Pauli and Dirac $R$-matrix approaches
is that calculations which require the resources of massively parallel supercomputer
centers can typically be carried-out on small local clusters. 
This facilitates the study of near neutral iron group ions which are omnipresent in 
cool astrophysical plasmas and which shine a light on their physical conditions.

\begin{acknowledgements}
The work of NRB was supported by the UK STFC under the UK APAP Network grant 
ST/J000892/1 to the University of Strathclyde.
The work of CPB was supported by US Department of Energy grants
to Auburn University. 
\end{acknowledgements}

\end{document}